\documentclass[9pt,twocolumn,twoside,lineno]{pnas-new}

\templatetype{pnasresearcharticle} 

\title{The advantage of leakage of essential metabolites and resultant symbiosis of diverse species}

\author[a]{Jumpei F. Yamagishi}
\author[b,1]{Nen Saito} 
\author[a,c,1]{Kunihiko Kaneko}

\affil[a]{College of Arts and Sciences, The University of Tokyo, 3-8-1 Komaba, Meguro-ku, Tokyo 153-8902, Japan}
\affil[b]{Graduate School of Science, The University of Tokyo, 7-3-1 Hongo, Bunkyo-ku, Tokyo 113-8656, Japan}
\affil[c]{Research Center for Complex Systems Biology, Universal Biology Institute, The University of Tokyo, 3-8-1 Komaba, Tokyo 153-8902, Japan}

\leadauthor{Yamagishi}

\significancestatement{
How can diverse species or strains coexist in microbial communities? Besides the fittest strain under isolation conditions, many others coexist. Population dynamics with appropriate cell-to-cell interaction would provide such diversity, but how this interaction is achieved remains elusive. Secretion of beneficial metabolites from the fittest strain could feed others and enable the coexistence; however, why do cells leak out these components essential to them? We numerically and analytically demonstrate that, counterintuitively, appropriate leakage of essential metabolites is beneficial to the leaking cells. A symbiotic relationship among diverse cells is then established: each cell cross-feeds others by secreting out essential metabolites that are usefully consumed by others, as if cells practice the competition in gift-giving called ``potlatch'' in human society.
}

\authorcontributions{Author contributions: J.F.Y., N.S., and K.K. designed the research; J.F.Y. conducted the research; J.F.Y., N.S., and K.K. analyzed the data; and J.F.Y., N.S., and K.K. wrote the paper.}
\authordeclaration{The authors declare that they have no conflicts of interest.}
\correspondingauthor{\textsuperscript{1}To whom correspondence should be addressed. E-mail: saito@ubi.s.u-tokyo.ac.jp or kaneko@complex.c.u-tokyo.ac.jp}


\begin{abstract}
Microbial communities display extreme diversity. A variety of strains or species coexist even when limited by a single resource. It has been argued that metabolite secretion creates new niches and facilitates such diversity. Nonetheless, it is still a controversial topic why cells secrete even essential metabolites so often; in fact, even under isolation conditions, microbial cells secrete \textcolor{black}{various} metabolites, including those essential for their growth. First, we demonstrate that \textcolor{black}{leaking} essential metabolites can be advantageous. If the intracellular chemical reactions include multibody reactions \textcolor{black}{like} catalytic reactions, this advantageous leakage of essential metabolites is possible and indeed typical for most metabolic networks via \textcolor{black}{``flux control'' and ``growth-dilution'' mechanisms; the later is a result of the balance between synthesis and growth-induced dilution with autocatalytic reactions.} Counterintuitively, the \textcolor{black}{mechanisms} can work even when the supplied resource is scarce. Next, when such cells are crowded, the presence of another cell type, which consumes the leaked chemicals is beneficial for both cell types, so that their coexistence enhances the growth of both. The latter part of the paper is devoted to the analysis of such unusual form of symbiosis: ``consumer'' cell types benefit from the uptake of metabolites secreted by ``leaker'' cell types, and such consumption reduces the concentration of metabolites accumulated in the environment; this environmental change enables further secretion from the leaker cell types. This situation leads to frequency-dependent coexistence of several cell types, as supported by extensive simulations. A new look at the diversity in a microbial ecosystem is thus presented.
\end{abstract}

\dates{This manuscript was compiled on \today}
\doi{\url{www.pnas.org/cgi/doi/10.1073/pnas.XXXXXXXXXX}}

\begin{document}

\maketitle
\thispagestyle{firststyle}
\ifthenelse{\boolean{shortarticle}}{\ifthenelse{\boolean{singlecolumn}}{\abscontentformatted}{\abscontent}}{}

\dropcap{I}n microbial communities, extremely diverse strains or species coexist \cite{Lozupone2012,Curtis2002,Datta2016}. Even when limited by a single nutrient, a variety of species coexist rather than a single fittest type competitively excludes all others \cite{Experiment1994,Goldford}. It has been argued that metabolite secretion can in principle create new niches and allow for their coexistence \cite{Zelezniak,Goldford,Maslov,arXiv1805}, whereas the competitive exclusion principle suggests that multiple species cannot coexist when growth is limited by the same single environmental resource \cite{Hardin,MacArthur1964}, known also as Gause's limit \cite{Gause}. Nonetheless, it is still unclear why cells secrete produced metabolites so often. 

Indeed, even under isolation conditions, microbial cells secrete various metabolites, despite the naïve expectation that leakage and loss of metabolites will hinder cellular volume growth.
Of course, it is evident that every cell should dispose of toxic compounds \cite{Wilkinson1974,inhibitory} or, according to classical syntrophy, inhibitory or waste byproducts \cite{Syntrophy,Syntrophy2,Syntrophy3}. 
Recent studies on the exometabolome \cite{ExometabolomicsMSI}, however, revealed that many microorganisms leak (and take up) a variety of metabolites that are necessary for growth, including most intermediates of central metabolism \cite{extendedOM}.  
Although some metabolic intermediates are considered possibly inhibitory at a substrate excess \cite{Pfeiffer-Bonhoeffer}, the leakage of various metabolites is observed even when the supplied resource is scarce \cite{extendedOM}. 
Why do cells secrete even essential metabolites so often? 
A simplistic answer is that small metabolites inevitably leak, regardless of whether the leakage inhibits cell growth. An alternative possibility, however, is that there are some benefits for cells when they leak chemicals necessary for their growth. Is such advantageous leakage really possible for a class of intracellular metabolic reactions, and if so, how is it possible and how general is it? These questions are addressed in the first part of this paper.

To answer \textcolor{black}{these} questions, we analytically and numerically investigated dynamical-system models of a cell with simple metabolic reactions. We show that the leakage of essential chemicals \textcolor{black}{(metabolites or enzymes)} can enhance cell growth even during nutrient limitation. 
\textcolor{black}{We found that} such advantageous leakage \textcolor{black}{becomes increasingly common as the number of components in metabolic reaction networks increases. One mechanism for this is flux control to increase the reactions leading to cell growth. We also unveil another general mechanism 
via global dilution due to the cell growth. In the latter mechanism, leakage of essential reactants for biomass synthesis can modify the negative feedback due to their dilution. With this dilution-synthesis balance, leakage of a reactant for biomass synthesis enhances the concentration of the other reactant(s), leading to an increase in cell growth.} 

In the first part, cell growth promotion by leakage of metabolites is considered for an isolated cell in a given chemical medium. On the other hand, the medium changes as metabolites are secreted by some cells, which will then affect the growth of other cells in the same medium. Now the second set of questions we address are at the level of a microbial community: how the cell--cell interactions mediated by this advantageous leakage influence the \textcolor{black}{communities} of cells of different types, and whether \textcolor{black}{they} can lead to \textcolor{black}{the} stable coexistence of diverse cell types (e.g., different species, strains, or mutants \cite{exometabolomics,Kashiwagi}), rather than the dominance of a single fittest type. 
\textcolor{black}{We will show that ``leaker'' and ``consumer'' cells (i.e., cells that beneficially leak some chemicals and that beneficially consume them) can immediately develop mutualistic relationships. 
Indeed, since the leaked chemicals are not waste but essential metabolites, cells of many other types can use them for their growth, whereas consumption of the leaked chemicals by other cells is beneficial for the leaker cells as the accumulation of the chemical in the medium is relaxed. Thus symbiotic relationships naturally and frequently develop.} 
As a consequence of each cell's optimization of its own growth, various types of cells actively leak and take up a variety of essential metabolites as if cells practice so-called ``potlatch,'' the ritual competition in gift-giving \cite{Mauss,Bataille}, eventually leading to symbiotic coexistence and prosperity of diverse cell types. This novel scenario will explain why the single strongest type does not dominate as a result of evolution.

\section*{\textcolor{black}{Formulation of Advantageous Leakage by an Isolated Cell}}
Let us consider an isolated cell that contains $n$ kinds of chemical components \textcolor{black}{(e.g., metabolites and enzymes)} as in Refs. \cite{FK1998,FK2012,Kaneko1994}. The cellular state is expressed by concentrations \textcolor{black}{(i.e., the number of molecules per the cellular volume $V$)} of the $n$ components, $\textbf{x}={}^t(x_0,x_1,\cdots,x_{n-1})$. In the cell, chemical $i$ is synthesized and decomposed by a set of intracellular reactions with rate $F_i(\textbf{x})$ and is exchanged with the environment at rate $f_i(\textbf{x};D_i,x_i^{(\mathrm{env})})$; if $f_i$ is positive, then chemical $i$ flows in from the environment, and if it is negative, chemical $i$ is leaked out. 
$D_i$ is a positive parameter characterizing the flow rate of each component $i$, \textcolor{black}{and we call it the diffusion coefficient. Note that the unit here is 1$/$time. 
The} fixed non-negative parameter $x_i^{(\mathrm{env})}$ represents the $i$th chemical's concentration in the environment; if chemical $i$ is an externally supplied nutrient, then $x_i^{(\mathrm{env})}$ is positive. Here we discuss the case of passive diffusion, where the flow rate of chemical $i$ is given by $f_i(\textbf{x};D_i,x_i^{(\mathrm{env})})=D_i(x_i^{(\mathrm{env})}-x_i)$, whereas the case of active transport is formulated similarly, with the uptake and secretion rates as $f_i(\textbf{x};D_i,x_i^{(\mathrm{env})})=D_ix_i^{(\mathrm{env})}$ and $=-D_ix_i$, respectively \cite{FK2012}. 

The time-dependent change in the $i$th chemical's concentration, $x_i$, can be written as 
\begin{eqnarray*}
\dot{x}_i=F_i(\textbf{x})+f_i(\textbf{x};D_i,x_i^{(\mathrm{env})})-\textcolor{black}{\mu(\textbf{x};\textbf{D},\textbf{x}^\mathrm{(env)})}x_i 
\end{eqnarray*}
where \textcolor{black}{$\mu(\textbf{x};\textbf{D},\textbf{x}^\mathrm{(env)})$ is the growth rate of the cell volume $V$ (i.e., $\mu\equiv \frac{1}{V}\frac{dV}{dt}$)}, and the third term represents the dilution of each chemical owing to the cellular volume \textcolor{black}{increase}. 
We assume that a steady state (i.e., a stable fixed point) $\textbf{x}=\textbf{x}^\ast$ exists and is reached, where $\textbf{x}^\ast$ satisfies $\textbf{G} (\textbf{x}^\ast;\textbf{D},\textbf{x}^{(\mathrm{env})})=\textbf{0}$ with $G_i(\textbf{x};\textbf{D},\textbf{x}^\mathrm{(env)})\equiv F_i(\textbf{x})+f_i(\textbf{x};D_i,x_i^{(\mathrm{env})})-\textcolor{black}{\mu(\textbf{x};\textbf{D},\textbf{x}^\mathrm{(env)})}x_i$. 
\textcolor{black}{In all the numerical simulations below and in the earlier studies \cite{FurusawaZipf2003}, cells reach a steady-growth state with constant concentrations after transient time. Further, non-oscillating growth is extensively observed in bacterial cells \cite{Hashimoto-Wakamoto2016}, whereas theoretical analysis based on the assumption of constant concentrations is consistent with transcriptome analysis of bacteria \cite{KKFurusawaPRX}. Note, however, even if oscillation states exist, we can just consider the time average of the concentrations and growth rate over an oscillation period or a cell cycle, and the results to be presented remain valid.} 

Now consider an infinitesimal change in diffusion coefficients: $\textbf{D} \to \textbf{D}+\delta\textbf{D}$, where $\delta D_i\geq 0$ if chemical $i$ is not \textcolor{black}{a} nutrient and $\delta D_i= 0$ otherwise. As long as chemical component $i$ is not externally supplied into the environment, an increase in the diffusion coefficient of \textcolor{black}{the} non-nutrient chemical $i$ leads to its additional leakage. Through this change, the steady state and growth rate also change as $\textbf{x}^\ast \to \textbf{x}^\ast+\delta\textbf{x}$ and $\mu^\ast\equiv\textcolor{black}{\mu(\textbf{x}^\ast;\textbf{D},\textbf{x}^\mathrm{(env)})} \to \mu^\ast+\delta\mu$. We consider infinitesimal $\delta \textbf{D}$ and analyze the value\textcolor{black}{s} of $\delta\textbf{x}$ and $\delta \mu$ by linearizing the equation $\dot{\textbf{x}}=\textbf{G}(\textbf{x};\textbf{D},\textbf{x}^\mathrm{(env)})$. 

By means of the Jacobi matrix $J=\partial \textbf{G}/\partial \textbf{x}|_{\textbf{x}=\textbf{x}^\ast}$, $\delta\textbf{x}$ and $\delta\mu$ are derived as follows (see SI Appendix for the derivation): \textcolor{black}{
\begin{eqnarray}
\delta\mu=\textcolor{black}{\left[\frac{\partial\mu}{\partial \textbf{D}}+\frac{\partial\mu}{\partial \textbf{x}}J^{-1}\left(\textbf{x}\frac{\partial \mu}{\partial \textbf{D}}-\frac{\partial \textbf{f}}{\partial \textbf{D}}\right)\right]}\cdot\delta\textbf{D}. 
\end{eqnarray}
}Because $\textbf{x}^\ast$ is a stable fixed point, all the eigenvalues of $J$ \textcolor{black}{have negative real parts.} Thus, the determinant of $J$ is nonzero, and the inverse matrix of $J$ exists. 

We will study how the infinitesimal leakage of a chemical can promote cell growth, via analytical and numerical calculation of $\delta\mu$. 
The results can be straightforwardly applied to multiple-chemical cases because the change due to leakage of multiple chemicals equals the sum of the changes due to leakage of each chemical if it is small.

We now investigate how $\delta\mu$ can be made positive by leakage of a necessary chemical, whereas leakage of  unnecessary chemicals is evidently advantageous, as seen in classical syntrophy \cite{Syntrophy,Syntrophy2,Syntrophy3} \textcolor{black}{(see also the example in Fig. S1). In this paper}, we assume that cell growth is determined only by the synthesis of biomass (or membrane) component(s) and that no chemicals directly retard the \textcolor{black}{biomass synthesis} because we are not concerned with unnecessary chemicals. We refer to \textcolor{black}{such useful} chemicals as ``leak-advantage'' chemicals when their leakage promotes the cell growth. 

\textcolor{black}{Before demonstrating the possibility of a leak advantage, we first note} that leakage cannot promote cell growth if intracellular chemical reactions include only one-body reactions like $i\to j$ (or, in general, $i\to j_1+j_2+\cdots+j_m$). Considering each elementary reaction, the additional leakage of substrate $i$ decreases the abundance levels of substrate $i$ and product $j$. Likewise, the leakage of product $j$ simply decreases its concentration without changing $x_i$. Because additional leakage of chemical $i$ or $j$ cannot increase their concentrations, in a system consisting of a combination of such one-body elementary reactions, the concentration of any chemical cannot be increased by leakage. Thus, leakage cannot increase the reaction rate of biomass synthesis, i.e., the growth rate of the system. This intuitive explanation is analytically proven in SI Appendix. 

On the other hand, if the intracellular metabolism includes multibody reactions \textcolor{black}{(e.g., catalytic reactions like $i+k\to j+k$)}, then the situation is different. 
\textcolor{black}{Indeed in the following subsections, we will discuss two possible basic mechanisms for the leak advantage to appear. We then demonstrate that the leakage of essential chemicals can often promote cell growth even for randomly chosen reaction networks consisting of metabolites and enzymes. In the following examples of reaction networks, $\textbf{x}^{(\mathrm{env})}$ is assumed to be $x_i^{\mathrm{(env)}}=S_\mathrm{env}$ if chemical $i$ is a nutrient, and $x_i^{\mathrm{(env)}}=0$ otherwise. \textcolor{black}{Also}, we assume passive diffusion, $f_i(\textbf{x};D_i,x_i^{(\mathrm{env})})=D_i(x_i^{(\mathrm{env})}-x_i)$; usage of active transport, however, does not alter the results as much.}
\begin{figure*}[ht]
\centering
\includegraphics[width=17.8cm]{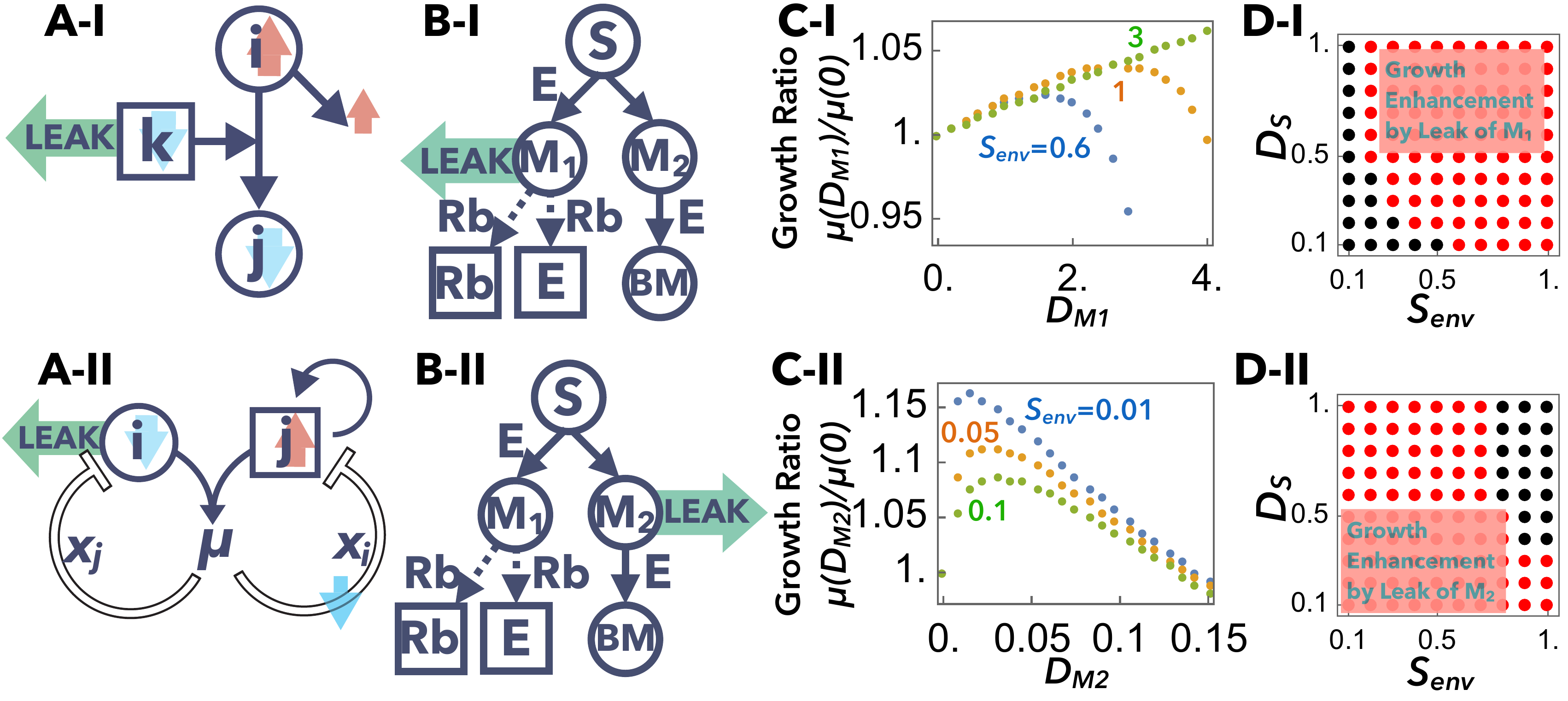}
\caption{
\textcolor{black}{The mechanisms underlying the leak advantage of an essential chemical: (I) metabolite $M_1$ or enzyme $E$ and (II) biomass precursor $M_2$. 
(A) Schematic illustration of the two mechanisms. (A-I) The flux control mechanism. Leakage of catalyst $k$ (or leakage of a metabolite, which causes an decrease in $x_\mathrm{k}$) decreases the flux of catalytic reaction $i+k\to j+k$, thereby decreasing the abundance of product $j$ but increasing that of substrate $i$. Consequently, it \textcolor{black}{may increase} the reaction rate of biomass synthesis (i.e., the growth rate) if $i$ also serves as a substrate or enzyme in another reaction. (A-II) The growth-dilution mechanism. Reactants ($i$ and $j$; $M_2$ and $E$ in the example of B-II) for the biomass synthesis have negative feedback with themselves; leaking $i$ relaxes the relative strength of this negative feedback for $j$, thereby increasing $x_j$. When the non-leaked chemical $j$ is involved in an autocatalytic module, the increase in $x_j$ exceeds the loss of $i$, and thus the biomass synthesis rate increases. 
(B) The reaction network to illustrate the two mechanisms. Solid and dashed arrows show catalytic reactions and translations, respectively. In this example, leakage of metabolite $M_1$ or enzyme $E$ (case I), or biomass precursor $M_2$ (case II) can enhance cell growth. Here, the growth rate $\mu$ is given by the synthesis rate of biomass $BM$, i.e., $\mu(\textbf{x})\equiv k_\mathrm{M_2\to BM}x_\mathrm{M_2}x_\mathrm{E}$. 
(C) The relationship between the diffusion coefficient $D_i$ and growth ratio $\mu(\textbf{x}(D_i))/\mu(\textbf{x}(D_i=0))$ (with different parameters). $D_\mathrm{S}$ is set to $1.0$. 
(C-I) The leaked chemical $i$ is metabolite $M_1$. Blue, orange, and green dots depict growth rates with $S_\mathrm{env}=0.6,1,3$, respectively. 
(C-II) The leaked chemical $i$ is metabolite $M_2$. Blue, orange, and green dots depict growth rates with $S_\mathrm{env}=0.01,0.05,0.1$. 
(D) Phase diagrams of the leak advantage with different rate constants. According to numerical simulations from Eq. [1], infinitesimal leakage of chemical $M_1$ or $E$ for (I) and $M_2$ for (II) is advantageous at $(S_\mathrm{env},D_\mathrm{S})$ of red dots. 
In (C-I) and (D-I), numerical simulations are conducted with rate constants $k_\mathrm{S\to M_1}=1$ and $k_\mathrm{M_2\to BM}=0.01$, while they are conducted with $k_\mathrm{S\to M_1}=0.9$ and $k_\mathrm{M_2\to BM}=0.25$ in (C-II) and (D-II). The other rate constants are set as $k_\mathrm{M_1\to Rb}=k_\mathrm{M_1\to E}=k_\mathrm{S\to M_2}=1$ in both cases.}
\label{fig:Matome}}
\end{figure*}

\subsection*{\textcolor{black}{Two Possible Mechanisms for the Leak Advantage with Simple Models for Illustration}}
\textcolor{black}{In this subsection, we consider two possible mechanisms for the leak advantage, ``flux control'' and ``growth-dilution'' mechanisms (schematically illustrated in Fig. \ref{fig:Matome}A). The former is a direct consequence of a leaked chemical enhancing the biomass synthesis, and the latter is a result of the balance between biomass synthesis and growth-induced dilution, together with autocatalytic nonlinear reactions.} 

\textcolor{black}{As a simple example, we consider the reaction network shown in Fig. \ref{fig:Matome}B, which consists of substrate $S$, enzyme $E$,
ribosome $Rb$, metabolites $M_1$ and $M_2$, biomass or biomembrane $BM$,
with the following chemical reactions and translations (solid and dashed arrows, respectively):
\begin{eqnarray*}
S+E\to M_1+E,\quad S\to M_2,\quad M_2+E\to BM+E,\\
M_1+Rb\dashrightarrow Rb+Rb,\quad M_1+Rb\dashrightarrow E+Rb.
\end{eqnarray*} 
Note that this reaction system is presented in order to exemplify the two general mechanisms: each ``chemical variable'' (and reaction) can also be interpreted as a cluster of molecules classified into categories, instead of a single specific molecule or elementary biochemical reaction.} 

The evolution of the concentrations is given by
\textcolor{black}{
\[\begin{cases}
\dot{x}_\mathrm{S}=-k_\mathrm{S\to M_1}x_\mathrm{S}x_\mathrm{E}-k_{S\to M_2}x_\mathrm{S}+D_\mathrm{S}(S_\mathrm{env}-x_\mathrm{S})-\mu(\textbf{x}) x_\mathrm{S}\\
\dot{x}_{\mathrm{M_1}}=k_\mathrm{S\to M_1}x_\mathrm{S}x_\mathrm{E}-(k_\mathrm{M_1\to Rb}+ k_\mathrm{M_1\to E})x_\mathrm{M_1}x_\mathrm{Rb}\\ \quad\quad\quad -D_{\mathrm{M_1}}x_\mathrm{M_1}-\mu(\textbf{x})x_\mathrm{M_1}\\
\dot{x}_{\mathrm{Rb}}=k_\mathrm{M_1\to Rb}x_\mathrm{M_1}x_\mathrm{Rb}-\mu(\textbf{x}) x_\mathrm{Rb}\\
\dot{x}_{\mathrm{E}}=k_\mathrm{M_1\to E}x_\mathrm{M_1}x_\mathrm{Rb}-D_{\mathrm{E}}x_\mathrm{E}-\mu(\textbf{x}) x_\mathrm{E}\\
\dot{x}_\mathrm{M_2}=k_{S\to M_2}x_\mathrm{S}-k_\mathrm{M_2\to BM}x_\mathrm{M_2}x_\mathrm{E}-D_{\mathrm{M_2}}x_{\mathrm{M_2}}-\mu(\textbf{x}) x_{\mathrm{M_2}}
\end{cases}\]
}where the growth rate is defined as the synthesis rate of biomass $BM$ from its precursor \textcolor{black}{$M_2$, so that $\mu(\textbf{x})\equiv k_\mathrm{M_2\to BM}x_\mathrm{M_2}x_\mathrm{E}$. 
Here, we do not consider the leakage of ribosome $Rb$.}

\textcolor{black}{The change in growth rate $\delta\mu$ due to leakage is obtained by numerically obtaining the steady state $\textbf{x}^\ast$ and Eq. [1]. The two mechanisms mentioned above are demonstrated by the two cases: (I) leakage of metabolite $M_1$ (or enzyme $E$) and (II) leakage of biomass precursor $M_2$. Although all chemicals---$S$, $E$, $Rb$, $M_1$, and $M_2$---are necessary for cell growth in this example, these leakages are advantageous over a wide range of parameters, as discussed below (Figs. \ref{fig:Matome}C, \ref{fig:Matome}D, and S2).} 
 
\textcolor{black}{{\bf Flux control mechanism}: this is schematically illustrated in Fig. \ref{fig:Matome}A-I, as provided by case (I). In this case, the leakage of $M_1$ (i.e., an increase in $D_{\mathrm{M_1}}$) decreases $x_\mathrm{Rb}$, the flux from $M_1$ to $E$, and $x_\mathrm{E}$. Accordingly, the flux $S+E\to M_1+E$ decreases, which raises $x_\mathrm{S}$. Then, the flux from $S$ to $M_2$ is upregulated (Figs. \ref{fig:Matome}C-I and S2A). In this way, the growth rate $\mu(\textbf{x})= k_\mathrm{M_2\to BM}x_{\mathrm{M_2}}x_\mathrm{E}$ increases if the rate constant for the reaction $S+E\to M_1+E$, $k_\mathrm{S\to M_1}$, is relatively large (Fig. S2B). Note that the essential factor of the present flux control mechanism is the decrease in $x_\mathrm{E}$. Thus the leakage of enzyme $E$ itself, instead of metabolite $M_1$, can also be advantageous, although leaking an enzyme may not be so common.} 

\textcolor{black}{This flux control mechanism may be straightforward, as the leakage directly alters the flux of chemical reactions. The other general mechanism based on the change in the dilution term, however, is indirect.}

\textcolor{black}{{\bf Growth-dilution mechanism}: this is schematically illustrated in Fig. \ref{fig:Matome}A-II, and by case (II) with the leakage of biomass precursor $M_2$. In this case, the dilution term due to the volume growth matters.} 

\textcolor{black}{Here, we call the reactants for biomass synthesis (i.e., biomass precursor $M_2$ and the enzyme $E$ for the biomass synthesis reaction in the example) ``biomass producers.'' 
Then, generally, every biomass producer has a negative feedback with itself: if the concentration of a biomass producer increases, then the rate of biomass synthesis and the dilution due to the volume growth increases, thereby suppressing its own concentration. 
For instance, the dilution term for $E$ in Fig. \ref{fig:Matome}B is $-\mu x_{E}=-k_\mathrm{M_2\to BM}x_\mathrm{M_2}x_\mathrm{E}^2$. Here, the magnitude of this negative feedback for a biomass producer (e.g., $E$ in the example) is weakened by decreasing the concentration(s) of the other biomass producer(s) (i.e., $x_\mathrm{M_2}$). The concentration of $E$ thus increases by leaking $M_2$. Due to the nonlinear autocatalytic processes for $E$ through $Rb$ and $M_1$, this increase can surpass the loss of the leaked biomass producer $M_2$; then the biomass synthesis rate, $k_\mathrm{M_2\to BM}x_\mathrm{M_2}x_\mathrm{E}$, is enhanced. Thus, the leakage of a biomass producer is (counterintuitively) advantageous. We call this mechanism the growth-dilution mechanism.} 

\textcolor{black}{One can check this analytically as follows. From $\dot{x}_{\mathrm{Rb}}=0$ and $\dot{x}_{\mathrm{E}}=0$, $x_\mathrm{Rb}=cx_\mathrm{E}$ and $k_\mathrm{M_1\to Rb}x_\mathrm{M_1}=\mu$ hold in the steady state, with $c\equiv\frac{k_\mathrm{M_1\to Rb}}{k_\mathrm{M_1\to E}}$; here we assume $D_\mathrm{E}=D_\mathrm{M_1}=0$ because only the leakage of $M_2$ is considered in this case (II), whereas even if they are positive, their values do not change the conclusion. From $\dot{x}_{\mathrm{M_1}}=0$, it follows 
\begin{eqnarray*}
x_\mathrm{M_1}=k_\mathrm{S\to M_1}\frac{x_\mathrm{S}x_\mathrm{E}}{\mu+c^\prime x_\mathrm{E}
}=\frac{1}{k_\mathrm{M_1\to Rb}}\mu,
\end{eqnarray*}
with $c^\prime\equiv c(k_\mathrm{M_1\to Rb}+ k_\mathrm{M_1\to E})$. 
Substituting $\mu(\textbf{x})=k_\mathrm{M_2\to BM}x_\mathrm{M_2}x_\mathrm{E}$ in the left-hand side of this equation, we gain
\begin{eqnarray*}
k_\mathrm{M_1\to Rb}k_\mathrm{S\to M_1}\frac{x_\mathrm{S}}{k_\mathrm{M_2\to BM}x_\mathrm{M_2}+c^\prime
}=\mu.
\end{eqnarray*}
Hence, the decrease in $x_\mathrm{M_2}$ by leaking $M_2$ decreases the denominator. Thus, the steady growth rate $\mu$ increases if the change in $x_\mathrm{S}$ is sufficiently smaller than the decrease in the denominator. This condition is satisfied when the reaction $S\to M_2$ is much faster than the reaction $S+E\to M_1+ E$, or when $k_\mathrm{M_2\to BM}$ is large enough (Figs. S2B and S2C).}

\textcolor{black}{Since the balance between biomass synthesis and growth-induced dilution determines the cellular steady state $\textbf{x}^\ast$, one can also use a self-consistent equation approach to calculate $\mu^\ast$ and $\delta\mu$ (see SI Appendix and Fig. S3 for details).}

\textcolor{black}{With the simple illustration so far, a requirement for both the mechanisms is suggested: 
the existence of autocatalytic module(s) (i.e., a positive feedback process to enhance its own reaction process and concentrations). One can see this property in Fig. \ref{fig:Matome}B. The synthesis of $E$ involves nonlinear autocatalytic processes as the precursor synthesis $M_1$ is catalyzed by $E$ and the synthesis of $E$ is catalyzed by $Rb$ synthesized from $M_1$.}

\textcolor{black}{For the flux control mechanism, positive feedback for the autocatalytic module may work excessively under certain conditions, due to the positive-feedback nature. 
In case (I), when $S_\mathrm{env}$ or the rate constant for the reaction into the autocatalytic module $k_\mathrm{S\to M_1}$ is large (Figs. \ref{fig:Matome}D-I and S2B) or rate constant $k_\mathrm{M_2\to BM}$ is small (Fig. S2C), this excessive production occurs and the leak advantage for $M_1$ (or $E$) appears.} 
 
\textcolor{black}{With regard to the growth-dilution mechanism, a non-linear autocatalytic process for $E$ is necessary to increase the growth rate more than the decrease by the leak of $M_{2}$. Note that this mechanism works even when the nutrient supply is scarce (i.e., $S_\mathrm{env}$ is small), as it is based on negative feedback via the growth-dilution mechanism. Indeed, the smaller the nutrient supply, the broader the parameter region for the leak advantage for $M_2$ (Figs. \ref{fig:Matome}C-II, \ref{fig:Matome}D-II, and S2). As long as the rate constant for biomass synthesis $k_\mathrm{M_2\to BM}$ is large or rate constant $k_\mathrm{S\to M_1}$ is small, the negative feedback due to the volume-growth dilution is relatively significant and thus this mechanism works (Figs. S2B and S2C).} 

\textcolor{black}{We here make an additional remark for case (I). In this case, the leakage of $M_1$ (i.e., an increase in $D_{\mathrm{M_1}}$) counterintuitively increases $x_\mathrm{M_1}$ when it enhances the cell growth; this is a consequence of nonlinear autocatalytic process involving ribosome and growth-dilution balance, as it is simply proven from the steady condition $\dot{x}_\mathrm{Rb}=0$ leading to $k_\mathrm{M_1\to Rb}x_\mathrm{M_1}=\mu+D_\mathrm{M_1}$.}

\textcolor{black}{ To close the subsection, we make some remarks on a few other examples. 
(a) For the simple example in this subsection, the volume growth is determined by the single biomass chemical $BM$; the first term of Eq. [1] is then zero and only the sign of its second term matters. However, even when all (or the leaked) chemicals contribute to the cell volume (the first term is then negative), the two mechanisms (especially, the growth-dilution mechanism) can work, as shown in Fig. S2D with the network of Fig. \ref{fig:Matome}B, in Fig. S4 with Example S-I, and in the next subsection. 
(b) For some networks like Example S-II in Fig. S5, the flux control mechanism can also work even when the nutrient supply is scarce. (c) A leak advantage is also possible even in a chain reaction system that does not include a chemical working as substrates for multiple reaction as studied above (e.g., see Example S-III in Fig. S6).}

\subsection*{Statistics of Randomly Generated Networks for Isolated Cell's Leakage}
\textcolor{black}{To examine if, and how commonly, the leakage of a beneficial chemical can promote cell growth, we randomly generated thousands of chemical reaction networks consisting of metabolites and enzymes. We considered reaction networks including only catalytic reactions like $i+k\to j+k$ with a catalyst $k$ and the equal rate constants (set at unity) as the simplest multibody reactions. 
Out of $n$ chemical components in each network, $N_\mathrm{enzyme}$ kinds of them are ``enzymes'' which can be the catalyst or product of each reaction, and a single nutrient and the rest of the chemicals (``metabolites'') can be the substrate or product of each reaction (see SI Appendix for details; Fig. S8 shows examples of randomly generated networks}). 
In the environmental condition fixed as $S_\mathrm{env}=0.1$ and $D_\mathrm{S}=1$, we checked whether the growth with each network is enhanced by increasing the diffusion coefficient, $D_i$, of each non-nutrient component $i$. 

\textcolor{black}{To explicitly include the cost of leakage, we here assume that all chemicals contribute equally to the cell volume: the growth rate is then defined as the gain rate of total components, $\mu(\textbf{x};\textbf{D},\textbf{x}^\mathrm{(env)})\equiv \sum_i f_i(\textbf{x};D_i,x_i^{(\mathrm{env})})$. In this case, the first term of Eq. [1] is always negative; that is, the leakage of non-nutrient chemicals, by itself, always decreases the cell volume. As shown below, the leak advantage can generally appear even in this case because the second term of Eq. [1] can be positive and surpass the first term.}

\begin{figure}[bt]
\centering \includegraphics[width=\linewidth]{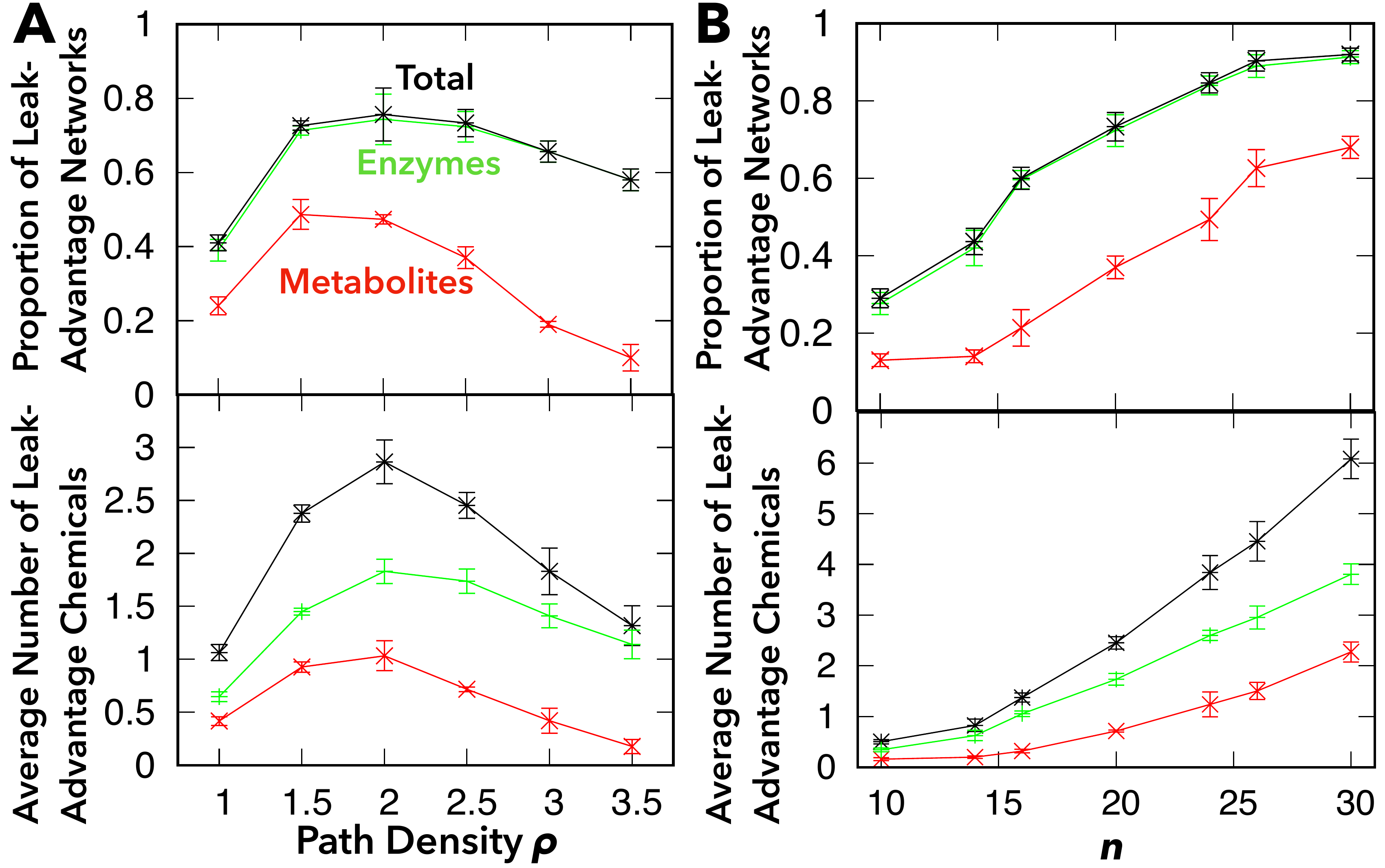}
\caption{Statistics of the leak advantage for randomly generated networks \textcolor{black}{in the model where all the chemicals equally contribute to the cellular volume and the growth rate is defined as $\mu(\textbf{x};\textbf{D},\textbf{x}^\mathrm{(env)})\equiv \sum_iD_i(x_i^\mathrm{(env)}-x_i)$. 
Three} hundred networks were randomly generated for each set of parameters. 
(A) Path density dependence of the proportion of leak-advantage networks \textcolor{black}{(top) and of the average number of leak-advantage chemicals (bottom)}. The number of chemicals $n$ is set to $20$. 
(B) The dependence on $n$ of the proportion of leak-advantage networks (top) and of the average number of leak-advantage chemicals (bottom). 
\textcolor{black}{Path density $\rho$ is set to $2.5$. In (A)--(B), red, green, and black lines show the value for the leakage of metabolites, enzymes, and chemicals in total, respectively. The error bars indicate one standard error, and the number of enzymes $N_\mathrm{enzyme}$ is set at $n/2-1$.} 
\label{fig:stat}}
\end{figure}

Figure \ref{fig:stat}A depicts the proportion of networks having a leak-advantage chemical \textcolor{black}{(either an enzyme or metabolite). Here, either an enzyme or a metabolite (or both) can leak, and the number of networks with leak-advantage metabolite(s), enzyme(s), or chemical(s) in total are} plotted as a function of path density $\rho$, where path density $\rho$ is defined as the number of all the reactions divided by the number of chemicals $n$, such that \textcolor{black}{the total number of catalytic reactions is $\rho n$.} 
Remarkably, the proportion of leak-advantage networks is greater than \textcolor{black}{50\% at $\rho= 1.5$ to $3.0$} in the case of $n=20$ (Fig. \ref{fig:stat}A), and Fig. \ref{fig:stat}B suggests that this proportion gradually increases with $n$. Hence, the presence of leak-advantage \textcolor{black}{metabolites and enzymes} seems to be a generic property of complicated catalytic reaction networks. 

Figure \ref{fig:stat}B also presents the average numbers of leak-advantage chemicals \textcolor{black}{and metabolites}, which also increase with $n$. \textcolor{black}{When $n\geq 16$,} each randomly generated network contains more than one leak-advantage chemical \textcolor{black}{(either enzyme or metabolite)} on average. Because metabolic networks in the cell contain a large number of chemical components, it is likely that leak-advantage chemicals are common. 

\textcolor{black}{Here, with the increase in $n$, there can be more autocatalytic modules of chemicals; this seems to be the reason why leak-advantage networks and chemicals are more common with larger $n$ in Fig. \ref{fig:stat}B. The peak in the proportion of leak-advantage networks at finite $\rho$ in Fig. \ref{fig:stat}A can also be understood: if path density $\rho$ is too low, each chemical is rarely involved in autocatalytic modules \textcolor{black}{nor multiple reactions}. On the other hand, if path density $\rho$ is too high, all the chemical reactions are extremely tangled, and the leakage of a chemical will reduce the flux of every reaction on average, and increasing specific reactions for biomass synthesis will be harder.} 

\textcolor{black}{The commonness of a leak advantage is also suggested by the results of other classes of models. The above results are qualitatively reproduced for a model which includes the ribosomal chemical and translation processes, and also for models in which the growth rate is determined by the synthesis rate of only a single biomass chemical  (Figs. S9 and S10).} 

\begin{figure*}[hbt]
\centering \includegraphics[width=\linewidth]{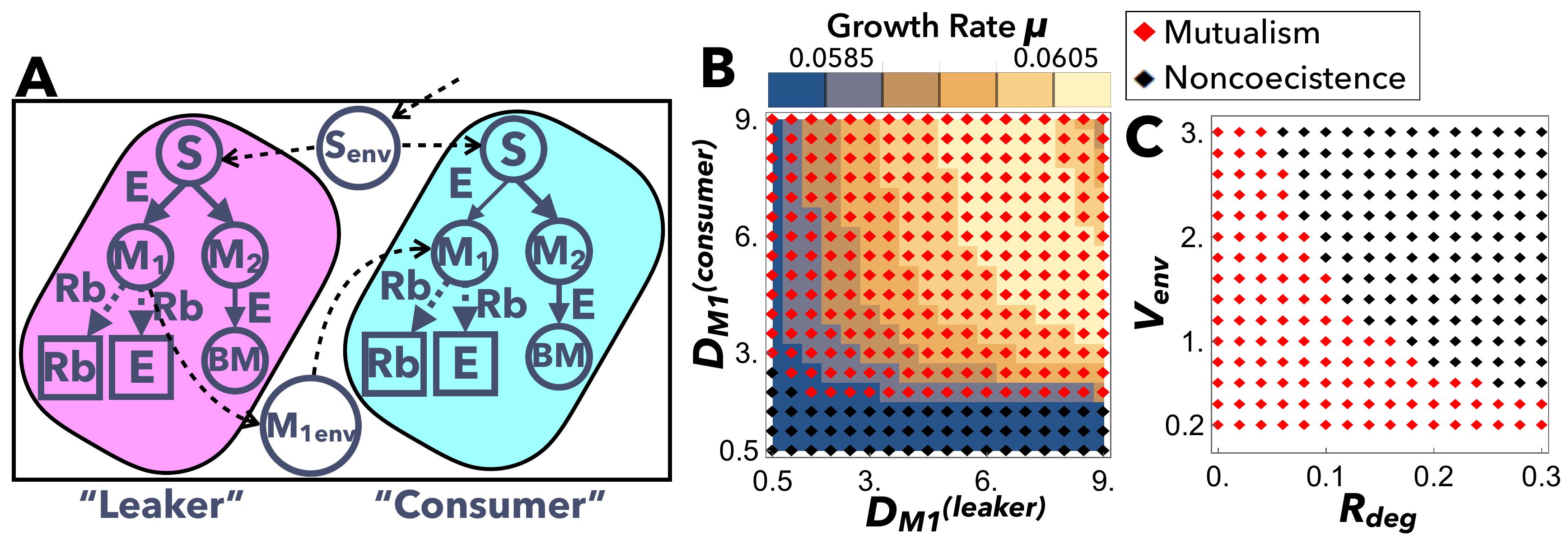}
\caption{An example of leaker--consumer mutualism: symbiosis between two cell types. 
(A) Schematic illustration of the mutualism between leaker (left) and consumer (right) cells. Both have the network structure  \textcolor{black}{of the example in Fig. \ref{fig:Matome}B}. 
(B) A phase diagram of symbiosis depending on \textcolor{black}{$D_\mathrm{M_1}^\mathrm{(leaker)}$ and $D_\mathrm{M_1}^\mathrm{(consumer)}$. The environmental parameters are set as follows: $V_\mathrm{env}=1$ and $R_\mathrm{deg}=0.1$.} The background color denotes growth rate $\mu$: a brighter color corresponds to higher $\mu$. 
(C) A phase diagram of symbiosis depending on environmental parameters, $R_\mathrm{deg}$ and $V_\mathrm{env}$. The diffusion coefficients of enzyme $E$ are fixed: $D_{\mathrm{M_1}}^\mathrm{(leaker)}=D_{\mathrm{M_1}}^\mathrm{(consumer)}=\textcolor{black}{2}$. 
In both (B) and (C), red and black diamonds represent mutualism and noncoexistence, respectively. \textcolor{black}{The rate constants are set as: $k_\mathrm{S\to M_1}^\mathrm{(leaker)}=1,k_\mathrm{S\to M_1}^\mathrm{(consumer)}=0.15,\; k_\mathrm{M_1\to Rb}^\mathrm{(leaker)}=k_\mathrm{M_1\to E}^\mathrm{(leaker)}=1,\; k_\mathrm{M_1\to Rb}^\mathrm{(consumer)}=k_\mathrm{M_1\to E}^\mathrm{(consumer)}=2,\; k_\mathrm{M_2\to BM}^\mathrm{(leaker)}=k_\mathrm{M_2\to BM}^\mathrm{(consumer)}=0.05$. The other parameters are set as $S_\mathrm{env}=D_\mathrm{S}=1$.} 
\label{fig:TWOsymbiosis}}
\end{figure*}

\section*{Symbiosis and Leakage in Ecology}
As discussed thus far, the growth rate of an isolated cell can be increased by leakage of some necessary chemicals. 
Given such advantageous leakage, cells will adopt a strategy to secrete metabolites into the environment to optimize their growth; this optimization may result from adaptation within a generation or evolution over generations. 
When cells of only one type that has a leak-advantage chemical are present, the secreted chemical \textcolor{black}{then} accumulates in the environment so that further secretion turns out to be harder. Then, if there are cells of a different type that benefit from consuming the secreted chemical, the concentration of the accumulated chemical diminishes, facilitating leakage from the former (secreting cells). On the other hand, such additional leakage also promotes the growth of the cells that consume the leaked chemical. 
This way, mutualism can be achieved via a secreted metabolite, and the growth rates of different cell types finally become equal. This state of affairs will make the coexistence of different strains or species possible, leading to the symbiosis of multiple cell types. 

To consider the cell--cell interaction by the transport of chemicals via the environment, 
external concentrations $\textbf{x}^\mathrm{(env)}$ are regarded as a variable. We then investigate the population dynamics of multiple cell types with different reaction networks and test whether they coexist in the common environment; the volume of the environment relative to the total volume of all the coexisting cells is designated as $V_\mathrm{env}$. 
The population fraction of cell type $j$, given by $p_j$, evolves according to the equation 
\begin{eqnarray}
\dot{p}_j=(\mu_j-\bar{\mu})p_j 
\end{eqnarray}
where $\mu_j$ is the growth rate of each cell type and $\bar{\mu}\equiv\sum_{j}p_j\mu_j$ is the averaged growth rate \cite{Kaneko2016}. 
In the external medium, the secreted components are weakly degraded or flowed out at rate $R_\mathrm{deg}$, so that the concentration changes as 
\begin{eqnarray*}
\dot{x}_i^\mathrm{(env)}=\sum_jp_jD_i^{(j)}(x_i^{(j)}-x_i^\mathrm{(env)})/V_\mathrm{env}-R_\mathrm{deg}x_i^\mathrm{(env)},
\end{eqnarray*}
if chemical $i$ is not a nutrient. If chemical $i$ is a nutrient, it is supplied into the environment via simple diffusion, so that the term $D_i^\mathrm{(env)}(S_\mathrm{env}-x_i^\mathrm{(env)})$ is added to the right-hand side of the above equation.

\subsection*{An Example of Leaker--Consumer Mutualism: Symbiosis between Two Cell Types}
We first consider the simplest situation: symbiosis between two cell types in which the leaker cells secrete a metabolite and the consumer cells take it up and consume it to grow. 

For simplicity's sake, the network structure of \textcolor{black}{the example in Fig. \ref{fig:Matome}B} is adopted both for the leaker and consumer cell types \textcolor{black}{(Fig. \ref{fig:TWOsymbiosis}A)}. The rate constants are different between the two, and chosen so that the leakage is beneficial only for the former type \textcolor{black}{and} the leaker's growth rate $\mu^\mathrm{(leaker)}$ with optimal diffusion coefficient \textcolor{black}{$D_\mathrm{M_1}^\mathrm{(leaker)}$} is higher than the consumer's growth rate $\mu^\mathrm{(consumer)}$ with $\textcolor{black}{D_\mathrm{M_1}^\mathrm{(consumer)}}=0$; otherwise, $\mu^\mathrm{(consumer)}$ is always greater than $\mu^\mathrm{(leaker)}$ and the consumer cell type is dominant in the environment. 

Numerical simulations showed that the mutualism between leaker and consumer cell types is actually achievable: with the diffusion coefficients corresponding to the red diamonds in Fig. \ref{fig:TWOsymbiosis}B, the leaker and consumer cells coexist (i.e., the growth rates of the two cell types are consistent) and the growth rate during coexistence is higher than \textcolor{black}{when} cells of only one type are present. Figure \ref{fig:TWOsymbiosis}B also indicates that the fastest growth is achieved by mutualism \textcolor{black}{between both} cell types. Accordingly, if both cell types adaptively alter their diffusion coefficients, mutualistic coexistence follows naturally. 

Figure \ref{fig:TWOsymbiosis}C reveals that the leaker--consumer symbiosis is achieved if $R_\mathrm{deg}$ and $V_\mathrm{env}$ are \textcolor{black}{small}, that is, if the secreted metabolite is efficiently transported to the other cell type. When degradation rate $R_\mathrm{deg}$ and environment size $V_\mathrm{env}$ are too large for \textcolor{black}{sufficient accumulation of the secreted metabolite} in the environment, the cell types no longer coexist, and only the leaker cell type survives. 

\begin{figure*}[bt]
\centering 
\includegraphics[width=17.8cm]{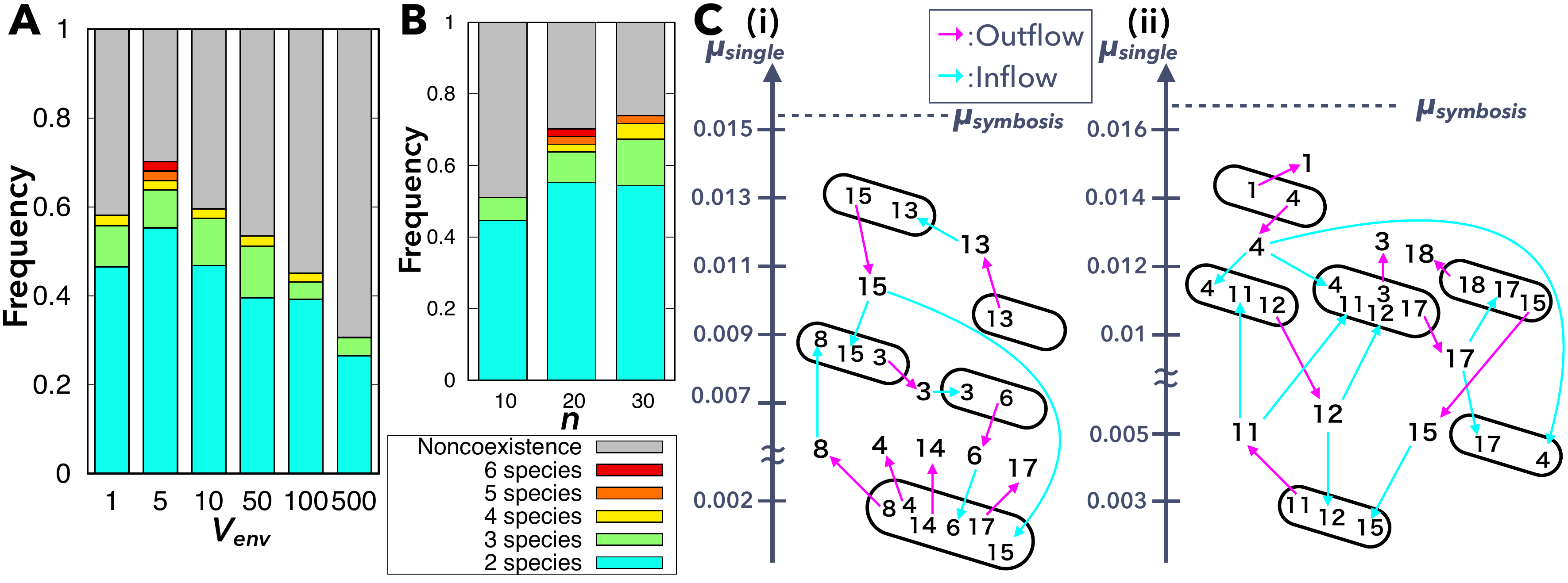}
\caption{Statistics of symbiosis among randomly generated networks. Approximately 50 independent trials were conducted for each set of parameters. 
(A) The dependence of the frequency of coexistence on $V_\mathrm{env}$. 
(B) The dependence of the population ratio on $n$. $V_\mathrm{env}$ is set to $5$. 
In both panels (A) and (B), the colored bars illustrate the frequency of symbiosis among two to six species, whereas the gray bar shows that of noncoexistence; the frequency is calculated from random implementations where the cell type with the fastest growth in isolation has a leak-advantage chemical in its reaction network. 
(C) Examples of metabolic exchange via the environment among five (i) and six (ii) coexisting cell types at $n=20$. Symbiosis among multiple cell types raises growth rate $\mu_\mathrm{symbiosis}$ higher than that in the case where a single cell type is present, $\mu_\mathrm{single}$. Pink and light-blue arrows respectively indicate leakage and uptake of each chemical component. 
In all the numerical simulations, the other parameters are fixed: $S_\mathrm{env}=0.1, D_\mathrm{S}^\mathrm{(env)}=10,D_\mathrm{S}=1,R_\mathrm{deg}=1\times10^{-5},\rho=2.5$. 
\label{fig:symbiosisDependency}}
\end{figure*}

\subsection*{Symbiosis among Randomly Generated Networks} 
In the last subsection, coexistence of multiple cell types via secreted metabolites can be stably achieved when each cell type changes its diffusion coefficients through adaptation. 
Indeed, the transport of chemicals between leaker and consumer cell types can be bidirectional if various chemicals are permeable, thus leading to more complicated forms of symbiosis. 

To investigate the possibility of symbiosis among more cell types, we extended the model of the last subsection to include a variety of cell types with different catalytic networks. 
New cell types with $N=50$ randomly generated networks are added into the environment one by one; then, the new cell type optimizes the diffusion coefficients so that its growth rate is maximal at environmental concentration $\textbf{x}^\mathrm{(env)}$. 
After the addition of each cell type, the population dynamics of Eq. [2] are computed over sufficiently long period $T$, until the population distribution reaches a steady state; here some (most) cell types may become extinct. 
After this procedure, each surviving cell type can gradually alter its diffusion coefficients so that its growth rate increases; all the coexisting cells simultaneously alter their diffusion coefficients until convergence (see SI Appendix for details).

\textcolor{black}{In this subsection, we adopted a model in which the synthesis rate of a single biomass chemical determines the cellular growth rate. For simplicity, metabolites are eliminated adiabatically as faster variables \cite{Kaneko1994}, so that every chemical can work as catalyst and substrate of some reactions. As shown in Fig. S10B, the leak advantage appears in the same way as in the reaction network model we studied in the first part.} 

The above model was numerically studied to test whether symbiosis among cells with randomly chosen different networks is achievable. Figure \ref{fig:symbiosisDependency}A illustrates the dependence of the proportion of samples manifesting symbiosis upon the size of the environment $V_\mathrm{env}$. As $V_\mathrm{env}$ is decreased, the cell--cell interaction \textcolor{black}{is} stronger because the secreted chemicals are less diluted. Hence $1/V_\mathrm{env}$ serves as an indicator of the strength of cell--cell interaction. Indeed, for smaller $V_\mathrm{env}$, symbiosis is achieved more frequently by exchange of metabolites via the environment. 
Note that when $V_\mathrm{env}$ is too small ($V_\mathrm{env}\simeq 1$, i.e., the total volume of cells equals that of the environment), the environmental concentration is sensitively affected by the addition of new cell types, and therefore coexistence of multiple cell types \textcolor{black}{turns} to be unstable. 

For $V_\mathrm{env}= 5$, symbiosis of multiple cell types in a single-nutrient condition is achieved in more than two-thirds of the trials as long as the \textcolor{black}{``strongest''} cell type (i.e., that with the highest growth rate in isolation) has a leak-advantage chemical (Fig. \ref{fig:symbiosisDependency}A). Figure \ref{fig:symbiosisDependency}B shows that the frequency of symbiosis tends to increase as the number of chemicals $n$ grows. This result suggests that symbiosis among multiple cell types via a leak advantage is commonly achievable for typical microbes that contain many chemical components. 

By examining how metabolites are exchanged via the environment (Fig. \ref{fig:symbiosisDependency}C), we demonstrate that coexistence of multiple cell types in a single-nutrient condition is achieved by the leakage and uptake of multiple metabolites. 
In an example of symbiosis among five cell types in Fig. \ref{fig:symbiosisDependency}C(i), the leaker--consumer relations via metabolic exchanges are hierarchical and cyclic: chemical $13$ unidirectionally flows into the strongest cell type from another type. Two cell types use the common leaked chemical ($15$) that leaks out from the strongest leaker cell type, whereas three cell types, including the above two, cyclically exchange different chemicals ($3,6,8$) to optimize their growth. Some chemicals ($4,14,17$) are leaked, but no cell types consume them. Another example of symbiosis among six cell types is shown in Fig. \ref{fig:symbiosisDependency}C(ii).

\section*{Discussion}
In this paper, we first demonstrated that microbial cells can optimize their growth by increasing the leakage of essential metabolites, rather than by changing enzymatic activity. The growth promotion by the leakage of essential components has been elucidated both by analytical and numerical analyses of models of simple intracellular reaction dynamics as well as of randomly chosen reaction networks. 
\textcolor{black}{One of the two basic mechanisms} behind this leak advantage is \textcolor{black}{flux control mechanism}: by reducing a flux in a multibody reaction path through passive or active secretion, another flux leading to cell growth is enhanced. 
\textcolor{black}{The other is growth-dilution mechanism: by modifying the negative feedback due to cell-growth dilution, the concentrations of some chemicals that directly contribute to biomass synthesis can be increased. These two mechanisms can work even under nutrient scarcity conditions (in contrast to the obvious expectation that metabolite secretion may be beneficial only with abundant nutrient \cite{Pfeiffer-Bonhoeffer,Hwa}), if the intracellular metabolism includes autocatalytic module(s) and chemical(s) involved in multiple reactions. Note that the existence of the autocatalytic process is naturally expected for a cell that grows exponentially in time. Especially, the synthesis of enzymes through translations by ribosome generally involves autocatalytic processes. For the flux control mechanism to work, the excess synthesis by a positive feedback should be weakened by leaking a component involved in the autocatalytic processes, as is intuitively understood. 
On the other hand, for the growth-dilution mechanism to work, some biomass producer(s) other than the leaked one must be involved in an autocatalytic module, such that the increase in their concentration(s) with the positive feedback exceeds the decrease in the leaked biomass producer. 
In particular, the later growth-dilution mechanism suggests specific candidates for chemicals giving rise to leak advantage: the precursors (such as amino acids), enzymes, and coenzymes that contribute to biomass synthesis.} 

\textcolor{black}{The leak advantage cannot appear if the intracellular metabolism includes no multibody reactions, as proven in SI Appendix. Although we have mainly considered catalytic reactions (the simplest multibody reactions), the above two mechanisms can work even with other kind of multibody reactions (e.g., Example S-IV in which a complex is formed as is adopted in Michaelis--Menten kinetics).}

In fact, many kinds of microorganisms secrete a variety of essential metabolites such as central-metabolic intermediates \cite{extendedOM} and vitamins \cite{vitamin} as well as hundreds of amino acids and sugars \cite{Baran2015,exometabolome,MetabolicNetwork}; an archaeon transfers lipids and possibly even ATP \cite{Syntrophy,TransferATP}. The present study suggests that the leakage of such metabolites indeed can be beneficial for cellular growth, and the control of leakage provides a possible means of adaptation. 
\textcolor{black}{Further, it may also explain why DNA, RNA, and exoenzymes such as proteases and nucleotidase are leaked \cite{exoenzymes} even though they also function within cells.} 
The present leak-advantage theory can be verified in bacterial and other microbial experiments by fixing the concentration of such secreted chemicals in the culture medium by means of a chemostat and by measuring the dependence of the cellular growth rate upon the extracellular concentration.

In the latter part of the paper, we showed that symbiosis among cells of different types can be achieved by employing the leak advantage. As the density 
of cells is increased, the metabolites secreted by leaker cells accumulate in the environment, 
thereby preventing further leakage. Even with active transport, metabolite accumulation causes higher costs for leakage because of the increase in the chemical potential \textcolor{black}{and} the cell growth is thus suppressed. Consequently, coexistence with a different cell type that consumes the leaked chemical for its growth is of benefit for the leaker cells, whereas the growth of the consumer cells is supported by the leaker cells. Both cell types increase their growth rates through cell--cell interactions mediated by the secreted metabolites. 
Indeed, facilitation of the growth by coexistence of different strains or species in several experiments has been reported \cite{exometabolome,ExperimentOfMetabolicCooperation,exometabolomics,FasterGrowth}. 
From a theoretical perspective, it should be noted that the coexistence of diverse cell types here is attained and analyzed only by adopting multilevel dynamics between intercellular population dynamics and intracellular metabolic dynamics and cannot be captured by \textcolor{black}{standard Lotka--Volterra type population dynamics.} 

In the leaker--consumer mutualism, the benefit for leaker cell types is indirect; it is due to the consumption of accumulated \textcolor{black}{beneficial} chemicals by consumer cell types, which is favorable when the density of leaker cells is high enough. 
The leaker--consumer mutualism is thus frequency dependent \textcolor{black}{and} whether it works depends on the degree of interaction via the secreted chemicals. In the present model, this degree depends on the relative volume of the medium toward that of a cell $V_\mathrm{env}$ (i.e., the inverse of cell density), and on the degradation rate of chemicals in the medium $R_\mathrm{deg}$. If $V_\mathrm{env}$ and $R_\mathrm{deg}$ are large enough, the leaker cells can continue to leak chemicals efficiently without the consumer cells, so that there is no room for synergy. In this sense, the leaker--consumer mutualism is different from ordinary forms of cooperation or division of labor \cite{JFY,cyanobacteria}. 
In some cases, however, each cell can simultaneously play roles of a leaker and consumer for different chemicals; metabolic division of labor is \textcolor{black}{then} achieved. 

\textcolor{black}{The origins of (and possible mechanisms allowing for) the microbial community of diverse cell types have often been discussed. A constructive laboratory experiment has revealed that cells with higher glutamine synthetase activity coexist with cells with lower activity, via leakage of glutamine synthesized by the former \cite{Kashiwagi}. Morris et al. have stressed the importance of chemical leakage by proposing the black queen hypothesis (BQH), a theory on the evolution of metabolic dependency based on gene loss \cite{LenskiBQH,MorrisBQH}. 
In these studies \cite{MorrisBQH,evoFBA,Zomorrodi2017}, however, whether the leakage is beneficial for the leaker cells is not fully addressed. Though the BQH also discusses the evolution of cooperation \cite{MorrisBQH,review1-5}, it assumes that, at the onset (i.e., before evolution proceeds), metabolic secretion leads to parasitism or free riding due to the properties of a permeable membrane. 
Although this assumption is not unreasonable, and is consistent with some empirical observations \cite{MorrisBQH,Gore2009,Nigel2011}, it is not clear why the cells have not evolved mechanisms to suppress the leakage (before cooperation via metabolic exchange evolves). In this respect, our results will strengthen the BQH:} some microbial cells secrete chemicals just because this process is beneficial for them. In this sense, the ``richer'' cells ``donate'' their products to ``poorer'' cells or dispose of these products, whereas this donation or disposal is also advantageous for the richer cells themselves, as if the cells are practicing a kind of ``potlatch'' often seen in human society \cite{Mauss,Bataille}. 

Indeed, the coexistence of multiple species via active secretion of chemicals has been discussed as classical syntrophy \cite{Syntrophy,Syntrophy2} in microbial communities, \textcolor{black}{where} it is generally assumed that the leaked chemicals are useless or inhibitory to the leaking species itself but are useful for the other species. Such chemicals could surely exist but, more frequently, the leaked chemicals are useful for both species. \textcolor{black}{By considering the latter case, the possibility of mutualistic coexistence is increased \textcolor{black}{and further, successive, mutual symbioses between multiple species can be established,} whereas the former case is not ruled out either.} 

Finally, let us discuss \textcolor{black}{whether} a leak advantage \textcolor{black}{is not eliminated in the course of evolution, by adopting appropriate gene-regulation of enzymatic activity, a well-known means to optimize cell growth \cite{Jacob-Monod}. 
If the leak advantage worked only for the cases to reduce some components in excess, then evolution in the metabolic network would eliminate such processes. In contrast, dilution by cell-volume growth is inevitable, and nonlinear autocatalytic processes are ubiquitous in cells. Thus, evolution to eliminate the growth-dilution mechanism would be more difficult. Moreover, elimination of the flux control mechanism through evolution would also not be so easy, as it can commonly appear in most networks.} 

\textcolor{black}{Instead of leakage, degradation by proteases could also reduce the abundance of proteins like enzymes. Leakage, on the other hand, is much simpler and does not need the synthesis of proteases which may be costly. Further, products of degradation such as amino acids (that can serve as precursors of biomass) can be disadvantageous, especially, for the growth-dilution mechanism to work.}

\textcolor{black}{In addition, evolutionary change in enzymatic activity to reach optimized growth without leakiness would be difficult, once the evolution progresses under the environmental conditions with interacting cells. As the cell numbers increase, the environment inevitably becomes crowded, and cell--cell interactions through secreted chemicals cannot be disregarded. In this case, optimization under isolation conditions no longer works, unless cells find an optimized solution without any secretion of chemicals. Once evolution with cell–cell interaction progresses, finding such a solution, even if it exists, would take many generations; before such isolated optimization is reached, other cell types that consume secreted chemicals will either emerge through evolution or invade from elsewhere, thereby also enhancing the growth of the leaker species. A symbiotic relationship with different cell types will then develop, as there are many such possibilities. Once this evolution of symbiosis with cell--cell interaction is established, it becomes increasingly difficult to find the ideal solution through evolution for a single species in an isolation condition that excludes all other species; rather, further complexification by additional species will evolve, as we saw in the last subsection.} 
Indeed, in some experiments, coexistence via metabolite secretion emerges \textit{de novo} \cite{exometabolome,SynEvo,Hom2014}, and nonspecific metabolic cross-feeding is reported to lead to coexistence of different phenotypes in such a community \cite{Goldford,Gordian}. 

\textcolor{black}{In summary}, we have shown that leakage of essential chemicals from cells can generally facilitate their growth, and such leaker cells can establish a symbiotic relationship with other cell types that use the leaked chemicals for their growth. This ``cellular potlatch'' generally emerges when the intracellular metabolic network is complex, which \textcolor{black}{provides} a basis for a complex microbial ecosystem with \textcolor{black}{diverse} strains.

\acknow{The authors would like to thank Chikara Furusawa for useful comments. This research was partially supported by a Grant-in-Aid for Scientific Research (S) (15H05746) and Grant-in-Aid for Scientific Research on Innovative Areas (17H06386) from the Ministry of Education, Culture, Sports, Science and Technology (MEXT) of Japan.}
\showacknow{}

\end{document}